# SRIB Submission to Interspeech 2021 DiCOVA Challenge

*Vishwanath Pratap Singh\*, Shashi Kumar\*, Ravi Shekhar Jha\*, and Abhishek Pandey*

Samsung Research and Development Institute Bangalore, India
vp.singh@samsung.com, sk.kumar@samsung.com, rs.jha@samsung.com, abhi3.pandey@samsung.com

## Abstract

The COVID-19 pandemic has resulted in more than 125 million infections and more than 2.7 million casualties. In this paper, we attempt to classify covid vs non-covid cough sounds using signal processing and deep learning methods. Air turbulence, vibration of tissues, movement of fluid through airways, opening and closure of glottis are some of the causes for production of acoustic sound signal during cough. Does the COVID-19 alter the acoustic characteristics of breath, cough, and speech sounds produced through the respiratory system? This is an open question waiting for answers.

In this paper we incorporated novel data augmentation methods for cough sound augmentation, and multiple deep neural network architectures and methods along with handcrafted features. Our proposed system gives 14% absolute improvement in area under the curve (AUC). The proposed system is developed as part of Interspeech 2021 special sessions and challenges viz. diagnosing of COVID-19 using acoustics (DiCOVA). Our proposed method secured $5^{th}$ position on leaderboard among 29 participants.

Index Terms: COVID-19, acoustics, machine learning, augmentation, respiratory diagnosis, healthcare [1]

## 1. Introduction

The COVID-19, novel coronavirus or SARS-Cov-2, has claimed thousands of lives and affected millions of people all around the world with the number of deaths and infections growing day-by-day. According to [1], the cough has highest prevalence among the symptoms of COVID-19 with 60.4% of all occurrence. For diagnosing COVID-19, the real-time polymerase chain reaction (RT-PCR) is a standard diagnostic test, but, it can be considered as a time-consuming test. Thus using the deep learning methods to classify non-covid and covid cough sounds can play a critical role in COVID-19 diagnosis.

Cough is a powerful natural respiratory defense mechanism that clears the central airways in human breathing system. The characteristics of the cough sound is dependent on the flow and configuration of the tissue elements involved, and as such is likely to change during cough as the physiological configuration evolves [2, 3]. The temporal pattern of the cough sound can be analyzed in three phases the first starts with an explosive burst, opening of glottis, followed by the second phase which is a period of noisy sound and slow decay of the noise as flow reduces due to glottal closure, transient forms the third phase [4]. These sounds can provide us enough information to distinguish between wet and dry cough, and ailment cough vs. voluntary cough [5]. Thus the ability to characterize the cough sounds of unhealthy speaker should therefore be helpful in diagnosis of the COVID-19.

Recently, several cough and breath data sets have been published. Examples include the 'Coughvid' [7], 'Breath for Science' [8], 'Coswara' [9], and 'CoughAgainstCovid'[10]. Multiple deep learning methods are proposed on these data sets to detect COVID-19 from the cough and breath sounds [11, 12, 13, 14, 15]. However, availability of limited training data is still a challenge in developing such systems.

In this paper, we experiment with different data augmentation methods, handcrafted features, small-footprint novel deep neural network architectures, and efficient model selection methods for evaluation. The system is developed as part of Interspeech 2021 special session and challenge on diagnosing COVID-19 using acoustics (DiCOVA) challenge. Our proposed method secured the $5^{th}$ position on leaderboard among 29 participants.

This paper is organised as follows. In Section 2, we explain the shared data and the overview of the challenge. In Section 3 and Section 4, we present the data augmentation methods and handcrafted features, respectively. In section 5, we discuss the model training methods, and deep neural network architectures. Results and analysis are presented in section 6. Conclusion with future works are provided in Section 7.

## 2. Data and Challenge Overview

The Shared data set contains 1040 audio files, stored in .FLAC format and a list containing train and validation lists for 5 folds. In each fold, the train set contains 822 audio files and remaining 218 audio files are part of validation set. Models need to be trained on all 5 folds without mixing the audio across the folds and need to be evaluated on respective validation set. Validation folds are used to select the best model across the folds. A separate blind test set containing 233 audio files is also shared for final evaluation on leaderboard. Shared data is part of Coswara data set [8]. It is to be noted that only 75 shared audio files are COVID-19 positive of which 50 audio files are part of train set and 25 are part of validation set. We also experimented 'Coughvid' [7] data set and obtained the results.

## 3. Feature Extraction Methods

First the raw cough speech signal is passed to energy threshold based speech activity detector and then normalized between -1 to 1. The spectrograms of cough sounds of speakers suffering from COVID-19 have been observed. There are marked differences observed in the spectrograms. It is noted that in non-covid cough sound higher proportion of signal energy lies

---

[1] * Represents the equal contribution

in higher frequency range compare COVID-19 cough sound signal as can be seen in Fig. 3 and Fig. 4.

We have experimented with 39 dimensional mel frequency cepstral coefficient (MFCC) and 24 dimensional handcrafted features to capture the time domain and frequency domain variability between COVID-19 negative and positive cough sounds. Handcrafted features include per frame energy of the signal(1D), fundamental frequency(1D), first four formants(4D), alpha ratio(1D) with cut-off frequency at 1400 Hz, relative average perturbation(1D), spectral flatness( D), kurtosis(1D), spectral contrast(7D), second order spectral polynomial(2D), spectral centroid(1D), spectral roll off(1D), and spectral bandwidth(1D), root mean square value of the signal(1D), and zero crossing rate of the signal(1D). These features are designed to capture the source and vocal tract variabilities. Then delta and double delta of the feature vector is computed to form 189 (3∗63) dimensional feature.

## 4. Data Augmentation Methods

### 4.1. Spectrum Interpolation

It is noted that only 75 cough audio files are COVID-19 positive and 50 out of them are part of train set in each fold. This makes the COVID negative vs COVID positive data ratio to 16:1 during training due to which the performance of classifier is suboptimal. We incorporate a novel spectrum interpolation method to increase the proportion COVID-19 cough sound sample in train set. First, we obtain the 1024 point discrete fourier transform (DFT) for each COVID-19 positive audio. Then, for each audio we obtain corresponding 5 nearest neighbour based on Euclidean distance of their DFT. Then, the linear combination of DFT of each COVID-19 positive audio and corresponding 5 nearest neighbour's is used to augment the COVID-19 cough sound. The linear combination coefficients are obtained from a uniform distribution between 0 and 1. In this way we obtain 5 augmented spectrum for each COVID-19 positive audio. We also experiment with the only nearest neighbour spectrum interpolation method to get 1 augmented spectrum for each COVID19 positive audio. Results for both of these methods are presented.

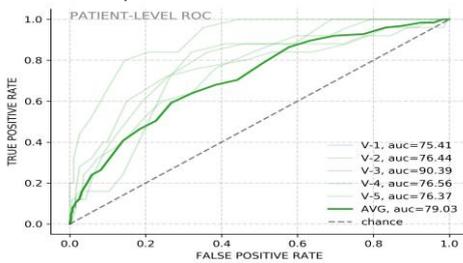

Figure 1: *Val ROC plot before data augmentation*

### 4.2. Noise Based Augmentation

Additive noise based augmentation has been proven to be beneficial for automatic speech recognition task [16]. Noise based augmentation improves the performance when the acoustic environment(background noise) are different during train and test time, which will be true in case system is to be used for diagnosis of COVID-19. We explore this method to augment the cough speech data set. The basic process of noisy training for deep neural network is as follows: first of all, sample some noise signals from some real-world recordings and then mix these noise signals with the original training data. Our noise set contains 1130 noise samples from kitchen, digital appliances, babble, music, traffic and other backgrounds. The signal to noise ratio (SNR) was varying between 5-20 dB and was uniformly distributed across the range. Noise based augmentation was performed for the entire data.

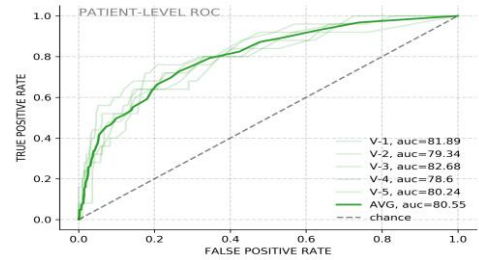

Figure 2: *Val ROC plot with data augmentation*

### 4.3. Vocal Tract Length Perturbation

Vocal Tract Length Perturbation (VTLP) is used in speech recognition to add the speaker to speaker variability that result primarily from differences in vocal tract length [17]. In this paper, we experiment with VTLP for cough sound augmentation. The warp factor was randomly selected between 0.85 to 1.15.

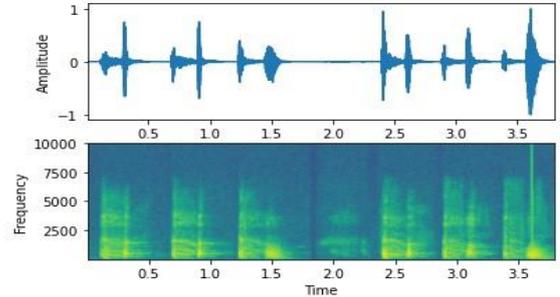

Figure 3: *COVID-19 positive cough sound signal and spectrogram.*

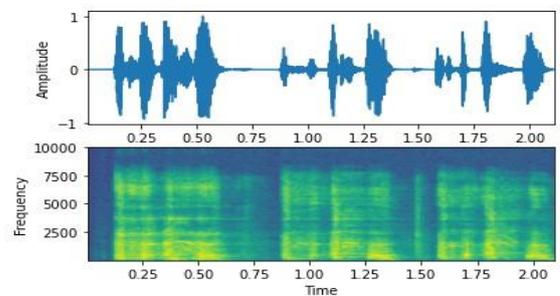

Figure 4: *Non-covid cough sound signal and spectrogram. It is noted that higher proportion of signal energy lies in higher frequency range compare COVID-19 cough sound signal.*

Table 1: *results of different models on combined validation and augmented validation set*

| model | fold | #layers | hidden dim | type | wtPos | norm | seq length | weight decay | AUC | specificity |
|---|---|---|---|---|---|---|---|---|---|---|
| LSTM | 1 | 1 | 48 | uni | 1 | utt-wise | 30 | 0 | 75.33 | 67.5 |
| LSTM | 1 | 1 | 48 | uni | 1 | utt-wise | 30 | 0.001 | 76.7 | 62.11 |
| LSTM | 2 | 1 | 64 | uni | 1 | utt-wise | 30 | 0 | 76.48 | 59.5 |
| LSTM | 2 | 1 | 64 | uni | 3 | utt-wise | 30 | 0 | 75.78 | 61 |
| LSTM | 1 | 1 | 64 | uni | 1 | utt-wise | 40 | 0 | 76.84 | 55 |
| LSTM | 2 | 1 | 48 | uni | 2 | utt-wise | 30 | 0 | 74.97 | 52.17 |
| LSTM | 1 | 1 | 48 | uni | 2 | utt-wise | 30 | 0 | 77.08 | 59 |
| LSTM | 1 | 1 | 48 | uni | 2 | utt-wise | 40 | 0 | 76.52 | 55.9 |
| LSTM | 5 | 1 | 48 | uni | 1 | utt-wise | 30 | 0 | 73.64 | 50.31 |
| LSTM | 1 | 1 | 128 | uni | 1 | utt-wise | 30 | 0 | 79 | 56.52 |
| LSTM | 2 | 2 | 48 | uni | 1 | utt-wise | 30 | 0 | 75.03 | 55.28 |
| LSTM | 3 | 2 | 64 | uni | 1 | utt-wise | 30 | 0 | 77.17 | 60.87 |
| LSTM | 1 | 1 | 64 | uni | 2 | utt-wise | 40 | 0.001 | 75.27 | 57.14 |
| LSTM | 3 | 1 | 48 | uni+5-fold | 2 | global | 40 | 0 | 75.47 | 54.04 |
| LSTM | 2 | 1 | 48 | seq-to-concat | 1 | utt-wise | 30 | 0 | 74.58 | 55.28 |
| LSTM | 4 | 1 | 48 | seq-to-concat | 1 | utt-wise | 30 | 0 | 74.89 | 48.45 |
| LSTM | 2 | 1 | 48 | seq-to-concat | 1 | utt-wise | 30 | 0.001 | 72.02 | 55.9 |
| LSTM | 3 | 1 | 48 | seq-to-last1 | 7 | utt-wise | 30 | 0.001 | 74.98 | 52.17 |
| LSTM | 5 | 1 | 48 | seq-to-last1 | 2 | global | 30 | 0 | 74.36 | 60.24 |
| LSTM | 4 | 1 | 48 | uni | 3 | utt-wise | 20 | 0 | 76.63 | 50.31 |
| LSTM | 1 | 1 | 48 | uni | 1 | global | 30 | 0 | 75.03 | 46.58 |
| LSTM | 1 | 1 | 48 | uni | 2 | global | 40 | 0 | 76.58 | 63.35 |
| LSTM | 4 | 1 | 48 | uni | 2 | global | 40 | 0 | 76.99 | 57.8 |
| LSTM | 1 | 1 | 48 | uni | 2 | global | 50 | 0 | 76.01 | 51.55 |
| LSTM | 1 | 1 | 48 | bidir | 2 | global | 40 | 0 | 76.07 | 63.35 |
| LSTM | 3 | 1 | 48 | bidir | 2 | global | 30 | 0 | 76.11 | 62.11 |
| LSTM | 1 | 1 | 48 | uni+coughvid | 2 | utt-wise | 30 | 0 | 76.72 | 67.08 |
| LSTM | 3 | 1 | 128 | uni+coughvid | 2 | utt-wise | 40 | 0.001 | 77.38 | 49.69 |
| LSTM | 1 | 1 | 32 | uni+AUROC loss | 1 | global | 30 | 0 | 77.54 | 69.57 |
| LSTM | 2 | 1 | 32 | uni+AUROC loss | 1 | global | 30 | 0 | 76.13 | 61.49 |
| ResNet-18 | 2 | NA | 32 | filter=3 | 1 | utt-wise | inCh=3 | 0 | 75.96 | 53.42 |
| ResNet-34 | 3 | NA | 32 | filter=3 | 1 | utt-wise | inCh=3 | 0 | 76.13 | 54.66 |
| CNN | 1 | 4 | 64 | filter=5 | 2 | utt-wise | inCh=3 | 0 | 75.96 | 59.63 |
| CNN | 1 | 4 | 64 | filter=5 | 1 | utt-wise | inCh=3 | 0 | 75.76 | 49.69 |
| CNN | 3 | 2 | 128 | filter=5 | 1 | utt-wise | inCh=3 | 0 | 75.46 | 56.28 |
| JVAE | 1 | 3 | 32 | uni | 2 | utt-wise | 30 | 0 | 74.29 | 60.25 |
| JVAE | 1 | 3 | 48 | uni | 2 | utt-wise | 30 | 0 | 74.11 | 50.31 |
| ensemble results on our set | | | | | | | | | 85.89 | 70.19 |
| ensemble results on blind test set (leaderboard) | | | | | | | | | 83.93 | 70.83 |

## 5. Classifier Description

We have experimented with standard machine learning algorithms such as, support vector machine (SVM) and random forest classifiers, and deep neural network architectures such as, LSTM, CNN and ResNet. We experiment with number of layers, sequence length and class weights in cross entropy (CE) loss.

We train LSTM based models in four different configuration. First type is regular uni-directional LSTM, termed as "uni". Second type is bi-directional LSTM, termed as "bidir". In model type termed as "seq-to-concat", we concatenate output vector of all elements in a sequence to form a super-vector. For example, if output dimension of LSTM layer is $n$ with a sequence length of $seqL$ then dimension of super-vector would be $n*seqL$. This super-vector is then passed through the final fully connected layer to predict probabilities. The "seq-to-concat" type model effectively processes $seqL$ number of frames before making predictions.

In "seq-to-last1" type model, output corresponding to the last timestamp of a sequence is passed further. This type of model makes prediction after processing all frames in a sequence. Here, the final fully connected layer takes *n* dimensional vector as input. It may be noted that both these model types, namely "seq-to-concat" and "seq-to-last1", make segment level predictions. All LSTM models use 20 frames as initial context before processing *seqL* number of frames for every sequence. The initial context frames set the hidden state of LSTM layer before processing actual inputs for which loss will be back propagated.

In CNN based models, we use 1-D convolution layers with filter length 5. Feature vector for each frame is reshaped into $63 \times 3$ before passing to CNN models implying that the input is now a 63 dimensional vector with 3 channels. After convolution layer, batch normalization is applied and then relu activation is applied. We also use Residual Networks [18], namely ResNet18 and ResNet-34, in our experimentation. To reduce learnable weights, we decrease filter size to 3 with 32 output channels for every convolution layer in default residual networks configuration.

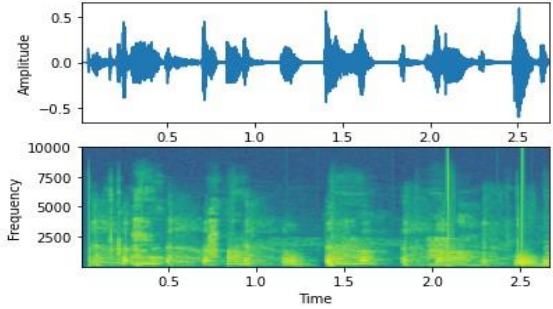

Figure 5: *Spectrum interpolation based augmentation of COVID-19 positive cough sound signal and spectrogram shown in Figure 1.*

5.1. AUROC Loss

In this challenge, our main task is to maximize area under curve (AUC). It is known that *CE* loss based classifier training does not have proportional relationship with AUC. We now formulate the AUROC loss function. We pick one non-covid cough sound at random and let $V_n$ be its predicted value. Similarly, we pick one covid cough sound at random and $V_p$ be its predicted value [19, 20, 21]. Then AUC can be seen proportional to the probability that the predicted values are in right order, that is $V_n < V_p$. It can be seen that this score is not differentiable as it does not make smooth transitions. So to force differentiability, we apply continuous approximation to this score using sigmoid function. Specifically, we calculate binary cross entropy (BCE) loss with assumpion that $Sigmoid(V_p - V_n)$ belongs to class 1. For stable model training, the final loss we use is given by

$$Loss = BCE(Sigmoid(V_p - V_n), 1) + BCE(Sigmoid(V_n - V_p), 0) \quad (1)$$

The model architecture we use with AUROC loss has only one output node. From dataset, we pick one non-covid cough sound sequentially and one covid cough sound randomly. We pass these two cough sounds through model and average the output of final fully connected layer for individual cough sounds. Thus, we have 1-D output value for each covid and non-covid cough. We then calculate the loss described in Eq 1 and back propagate the gradients to train the model.

5.2. JVAE

Following [22], the variational lower bound (ELBO) of joint VAE (JVAE) can be written as

$$\begin{aligned}\mathcal{L}_2(\theta, \phi; x, \boldsymbol{y}) &= \int_z q_\phi(z|x, \boldsymbol{y}) \log \frac{p_\theta(x, \boldsymbol{y}, z)}{q_\phi(z|x, \boldsymbol{y})} \\ &= E_{q_\phi(z|x)} \log p_\theta(x|z) + E_{q_\phi(z|x)} \log p_\theta(y|x, z) \\ &\quad - KL(q_\phi(z|x) \| p_\theta(z)) \quad (2)\end{aligned}$$

To train JVAE model, we minimize negative of the ELBO described in Eq 2. In original JVAE formulation, all the above conditional distributions are modelled by diagonal Gaussian distribution. We now propose to model $p_\theta(y|x,z)$ by a binomial distribution. Now, minimizing negative of $E_{q_\varphi(z|x)} \log p_\theta(y|x,z)$ in ELBO (Eq 2) reduces it to binary cross entropy (BCE) loss. In practice, the actual loss used to train JVAE network is given by

$$Loss = \lambda_1 MSE_x + \lambda_2 BCE_y + \lambda_3 KLD, \quad (3)$$

We choose value of hyper parameters in Eq 3 following [22]. Thus, $\lambda_1$ is taken as 1, $\lambda_2$ is taken as 10 and $\lambda_3$ is taken as 0.1.

## 6. Results

Patient level receiver operating characteristic (ROC) on validation set is presented in Fig. 1 and Fig. 2. It can be observed that the proposed data augmentation methods and architectures give significant improvement in the area under the curve (AUC). Detailed results are shown in Table 1. The first column depicts the type of model used. The "fold" column mentions the dataset fold on which the model is trained and validated. For all the experiments, we train our classifier model on each of the 5 folds of dataset and then choose the best model. Hidden dimension of layer before the final fully connected layer in our model architecture is presented in "hidden dim" column. Variation in model architecture type and its specialities is disclosed in "type" column. Details of these variations are discussed in Section 5.

In *CE* loss, values in "wtPos" column is multiplied with loss corresponding to covid positive cough data. This is done to counter the imbalance of covid vs non-covid cough sounds in dataset. We normalize the input feature vector with mean before passing through classifier model during both training and testing. The "utt-wise" token in "norm" column means mean is calculated per-utterance. We also calculate mean of all dataset beforehand, termed as global-mean. The "global" token in "norm" column specifies that the global-mean vector is used for feature normalization. This is done to ensure that handcrafted features do not vanish for any input cough sound. It also ensures that inter-frame variations in handcrafted features and their absolute values remain expressive.

The sequence length parameter of LSTM models is reported in "seq length" column. We experiment with sequence lengths 20, 30, 40 and 50. For CNN and residual networks, batches are not prepared sequentially. So, we mention number of channels ("inCh") of input features. The Adam optimizer in pytorch toolkit offers weight regularization in form of "weight decay". We use weight decay with a factor of $10^{-3}$ for training of some models. Rest of the models are trained without weight decay – mentioned as 0 in "weight decay" column. We report AUC yielded by classifier models in "AUC" column. We also specificity at > 80% sensitivity in "specificity" column of Table 1.

## 7. Conclusion

In this paper we explore various deep learning architectures, data augmentation techniques, and feature extraction method to improve state of the art cough sound based COVID-19 detection system. Our proposed system gives 14% absolute improvement in area under the curve (AUC) compare to baseline system released as part of the challenge.

Future work includes the experimentations with attention based architectures on raw audio. It will also be interesting to explore the developed method for other cough symptom based ailments.